\documentclass[10pt]{article}
\usepackage[utf8]{inputenc}
\usepackage[pdftex]{graphicx}
\graphicspath{{./pdf/}}

\usepackage[dvipsnames]{xcolor}
\usepackage{hhline}
\usepackage{amssymb}
\usepackage{hyperref}

\definecolor{codeblue}{rgb}{0.25,0.5,0.75}
\definecolor{backgray}{rgb}{0.95,0.95,0.95}

\usepackage{listings}
\lstdefinestyle{pythonstyle}{
	language=Python,                 
	backgroundcolor=\color{backgray}, 
	commentstyle=\color{OliveGreen},       
	keywordstyle=\color{blue},        
	stringstyle=\color{codeblue},     
	basicstyle=\ttfamily\footnotesize,
	frame=single,                     
	breaklines=true,                  
	numbers=left,                     
	numberstyle=\tiny\color{gray},    
	tabsize=4,                        
	showstringspaces=false,            
	morekeywords = {QMLModel, ModelFinder}
}
\newcommand{\gcm}{{\color{OliveGreen} \checkmark}}
\newcommand{\rx}{{\color{red} $\times$}}

\title{AQMLator --- An Auto Quantum Machine Learning E-Platform}

\author{Tomasz Rybotycki \and Piotr Gawron}
\date{}

\begin{document}
	
\maketitle

\begin{abstract}

	A successful Machine Learning (ML) model implementation requires three
	main components: training dataset, suitable model architecture and training
	procedure. Given dataset and task, finding an appropriate model might be
	challenging. AutoML, a branch of ML, focuses on automatic architecture
	search --- a meta method that aims at removing human from ML system design
	process. The success of ML and the development of quantum computing (QC) in
	recent years led to a birth of new fascinating field called Quantum Machine
	Learning (QML) that, amongst others, incorporates quantum computers into ML
	models. In this paper we present AQMLator, an Auto Quantum Machine Learning
	platform that aims to automatically propose and train the quantum layers of
	an ML model with minimal input from the user. This way, data scientists can
	bypass the entry barrier for QC and use QML. AQMLator uses standard ML
	libraries, making it easy to introduce into existing ML pipelines.
	
\end{abstract}


\sloppy

\section{Introduction}
	Machine learning (ML) is one of the fastest-progressing research directions
	in applied computer science. The field investigates the development of
	algorithms that can learn from data by fitting a collection of
	model parameters to the data via iterative optimization of an objective
	function. The selection of a model structure, be it a neural network
	or kernel function, is a problem-dependent task often made by
	hand. But there exist Auto ML systems~\cite{AutoML2019} that can choose a
	model automatically depending solely on the input data and the task at
	hand.

	Quantum computing (QC) studies how hard computational problems can be
	efficiently solved using quantum mechanics. A large-scale
	error-corrected quantum computer can solve computational problems that
	don't have a classical solution. A prime example of that is Shor's
	algorithm~\cite{Shor} for integer factorization. The ``holy grail'' of
	applied QC is the so-called quantum supremacy or quantum advantage. That is
	the name for a technological milestone marking the moment when
	quantum machines will solve a specific task faster than the most advanced
	supercomputer. Although there have already been several quantum supremacy
	claims in recent years~\cite{arute2019quantum}, there are yet no practical
	problems solvable only by using quantum computing.

	The search for such practical problems focuses on applications in the soft
	computing areas that are less susceptible to the current quantum hardware
	imperfections. One of the possible applications of QC is quantum machine
	learning (QML)~\cite{biamonte2017quantum}. This field of science investigates how quantum computers can be employed
	to build ML models that can be fit to data and then used during the
	inference process. In one of the QML scenarios, a variational quantum
	circuit forming a quantum neural network (QNN) constitutes only one part of
	the ML data processing pipeline. Since designing such a pipeline with a
	quantum component is challenging for non-experts in QC, we propose an auto
	ML solution that suggests ready-to-use QML models.
	
	This paper is organized as follows. In the next section, we present a short
	overview of the state-of-the-art. We point out the challenges laid before
	auto(mated) (quantum) machine learning, the most recent techniques tackling
	these problems and the related software. In section~\ref{sec:aqmlator}, main
	of this work, we present AQMLator, an auto quantum machine learning (AQML)
	e-platform designed to fill the gap in the AQML literature. We overview its
	construction, design principles and key features. The paper is concluded
	with some insights about future works.

\section{State of the art}

	As machine learning is getting progressively more complex, a need for
	methods of its automation arises. Complicated ML tasks like neural
	architecture search (NAS), hyperparameters optimization (HPO) or even model
	selection (MS) are often beyond the abilities of a scientist from outside
	the field. Automated machine learning was introduced as a way to address
	these shortcomings~\cite{Hutter2019}. 
	
	With the emergence of a new interdisciplinary field like quantum machine
	learning, a revision of the AutoML is required. QML models are often vastly
	different from their classical counterparts. While several standard
	techniques of tackling typical ML model creation problems still apply,
	automated quantum machine learning may require some adjustments due to the
	quantum	nature of its models. For instance, when discussing QML, one no
	longer talk about NAS, but rather QAS (Quantum Architecture Search). A
	recent, extensive summary of differences between quantum and classical ML
	can be found in~\cite{Cerezo2022}. Adding quantum computing to the list of
	skills required to properly introduce a QML model to one's research, it's
	clear why the non-experts may be discouraged from using such models. That's
	exactly the	reason why Auto(mated) Quantum Machine Learning is required. 

	As pointed out in~\cite{Cerezo2022}, one of the crucial challenges in QML is
	development of efficient quantum architecture search techniques. Given that
	parametrized (variational) quantum circuits can also be viewed as quantum
	machine learning models~\cite{Benedetti_2019} QAS has significant impact on
	the whole field of quantum computing. That's especially true if we
	additionally consider low circuit depth requirement of the current,
	NISQ-era\footnote{Noisy intermediate scale quantum.} machines.
	
	There are many proposition for quantum architecture search automation. A
	vast area of these research is occupied by reinforcement learning-based
	techniques. They use an ML agent to autonomously explore the search space
	of possible circuits in order to find the architecture optimal for a given
	task~\cite{ostaszewski2021, kunduPhd}. Recent works on QAS also use
	Kolmogorov-Arnold Networks that has gained some recognition in recent
	months~\cite{kundu2024}. These techniques, and a plethora of others
	(see~\cite{martyniuk2024survey}), although autonomous to some degree, use
	classical ML to find the optimal quantum circuit architecture. In order to
	use, and especially tweak them, specialized knowledge is required.
	
	In a typical scenario, before one even tries solving N/QAS, they first
	consider what class of (Q)ML models they should use. With a development as
	rapid as in the case of ML, it comes as no surprise that this itself is a
	daunting task. Even if we restrict the their selection to hybrid-QML
	models, the number is still substantial~\cite{bowles2024}. Considering
	how vital model selection is for the research~\cite{raschka2020}, how
	expensive the training of a model can be~\cite{Guerra2023} and how the
	access to quantum resources is currently limited, a need of guidance in
	that matter becomes clear. Although~\cite{Gentile2021} addresses this issue
	to some degree, the work is very technical and requires QC know-how. 
	
	When it comes to hyperparameter optimization, the situation is significantly
	better. Their importance~\cite{Moussa2022, Moussa2024} and optimization
	\cite{herbst2024} is a topic of multiple studies, especially in the context
	of quantum neural networks. These studies, although insightful, are often
	technical and don't	provide any software. On the other hand, classical
	approaches have been shown to work well for QML models.
	In~\cite{lusnig2024}, for instance, the authors use Optuna hyperparameter
	optimization framework~\cite{optuna} to find hyperparameters for their
	hybrid models. What's especially important in the context of this work is
	that Optuna doesn't require any knowledge about its underlying algorithms,
	and thus can be considered an automated hyperparameter optimizer.
	

	All the above solutions, albeit addressing the issues expected from
	automated machine learning system, never explicitly claimed to be such
	system. There's however a number of works where the authors regard their
	solutions as such. It's vital to start with a recent
	whitepaper~\cite{klau2023}. Its authors review current classical AutoML
	solutions and their possible extensions for quantum models. In their other
	work~\cite{klau2023qml}, some of the authors present the AutoQML, an actual
	Python auto quantum machine learning framework. In~\cite{ALTARESLOPEZ2024}
	the authors propose other so-called AutoQML method that use genetic
	algorithms for automatic generation of trained, quantum-inspired
	classifiers. The term quantum-inspired was meant to underline the fact
	that the resultant model was supposed to run on a fully classical hardware
	despite it using unitary gates --- common building block in quantum
	computing. Given the nature of this approach, it seems to be applicable for
	QML models. In~\cite{subasi2023} the authors propose a multi-locality search algorithm for initial points, circuit parameters and data
	preprocessing options. This paper, however, is highly technical. Finally,
	there are also signs of practical use of AQML for Wi-Fi sensing
	task~\cite{koikeakino2022}, where the authors claim to use their so-called
	AutoAnsatz approach for quantum architecture search and hyperparameter
	optimization. We found that the code was provided only for the genetic-oriented AutoQML \cite{Altares-Lopez2023}, even though the first AutoQML team has a GitHub repository~\cite{AutoQML_repo}.

\section{AQMLator}
\label{sec:aqmlator}

	We call our solution \textbf{A}uto \textbf{Q}uantum \textbf{M}achine
	\textbf{L}earning platform --- AQMLator~\cite{AQMLatorPyPI}. It aims to
	enable data scientists, especially those without knowledge of QC, to use QML
	models. Given a task and the data, AQMLator will seek the model that
	fits it best. As a result, the platform will propose a quantum model, its
	circuit structure and its weights. The user can then use the suggested
	model as-is or encompass it in a hybrid quantum--classical model e.g. by
	extracting the proposed model as a torch
	layer~\cite{paszke2017automatic}.

\begin{figure}[htbp] 
		\centering
		\includegraphics[width=0.9\textwidth]{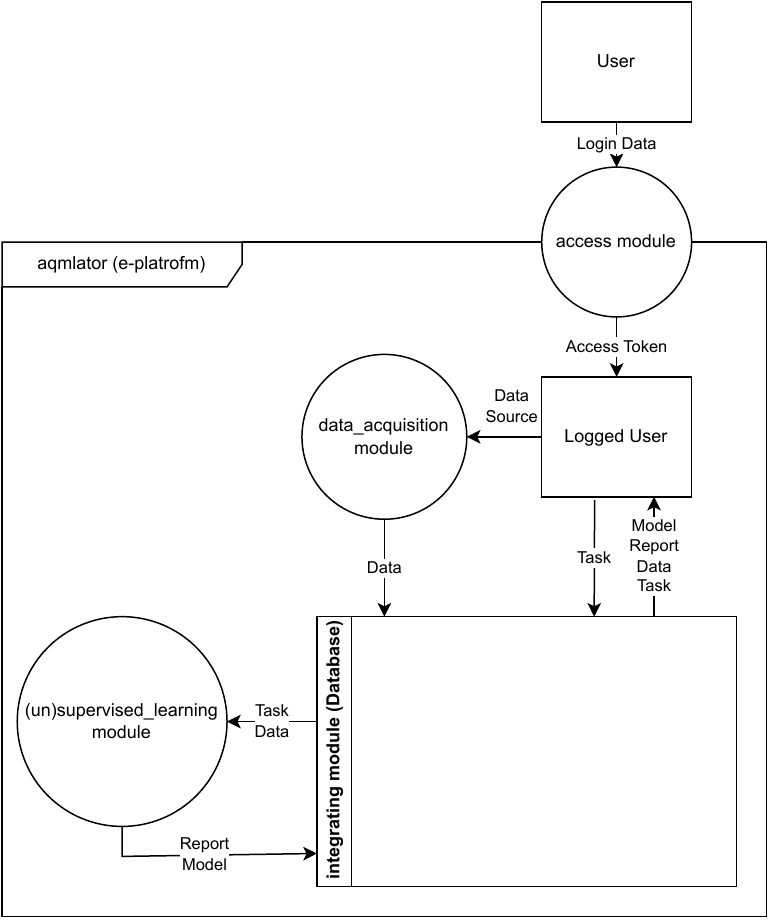}
		\caption{Data flow diagram of the AQMLator platform. 
			Notice that the user's input is reduced only to task and data source
			specification.}
		\label{fig:data_flow_diagram}
	\end{figure}

	In Figure~\ref{fig:data_flow_diagram}, we present the data flow diagram of
	the AQMLator platform. It shows that the only input required from the user
	is the source of data and the task the user wants the model to perform. No
	specialized knowledge is required.
	
\begin{figure}[htbp]
		\centering
		\includegraphics[width=\textwidth]{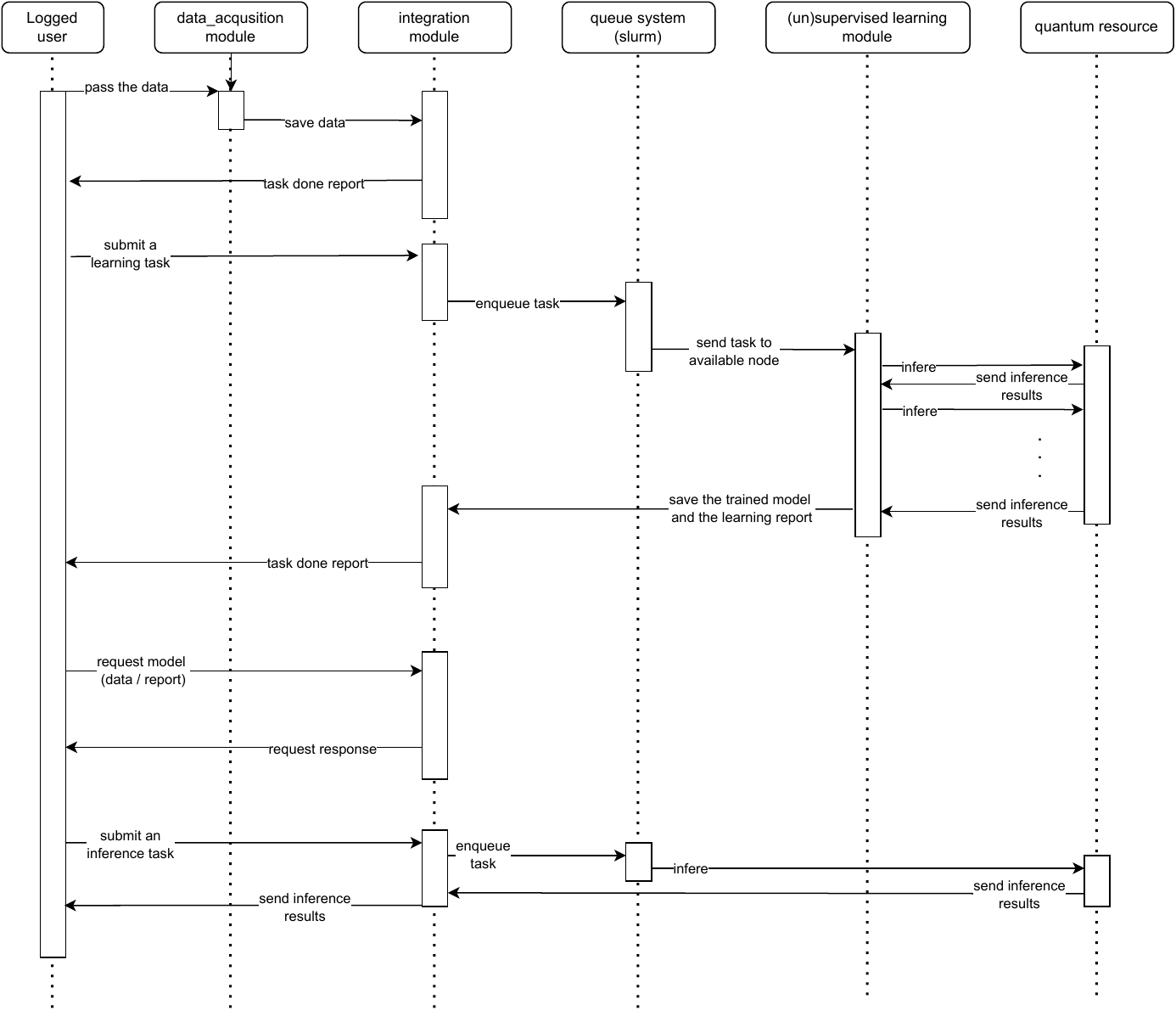}
		\caption{Sequence diagram of the quantum model training and
			inference process in AQMLator. The quantum resource we
			use can be classical (simulator) or quantum and, in general, can
			vary between the runs.}
		\label{fig:sequence_diagram}
	\end{figure}

	AQMLator can be used with simulators or physical quantum devices. In
	Figure~\ref{fig:sequence_diagram}, we present the sequence diagram of the
	model training and inference process using the our platform. The diagram
	shows that the device we use during the training may differ from the one
	used for inference. We can also see that throughout the training, AQMLator
	strives to minimize the reservation time and the number of costly quantum
	device calls. In that sense it is quantum resource aware (QRA).

	As the goal of AQMLator is to provide a complex auto quantum machine
	learning solution, it had to address all of the challenges mentioned in
	the section above. It does not only that, but also provides the source
	code, abiding to the open science practices. In Table~\ref{tab:comparison}
	we summarize features offered by the current auto quantum machine learning
	systems. 

	\begin{table}[!htbp]
		\centering
		\caption{Summary of the key features offered by the state-of-the-art
		AQML solutions. Notice that only AQMLator satisfies all of them. Features not mentioned in the respective works were marked with \rx.}
		\label{tab:comparison}
		\begin{tabular}{|c|c|c|c|c|c|}
			\hline
			\textbf{Solution} & \textbf{MS} & \textbf{QAS} &  \textbf{HPO} &
			\textbf{QRA} & \textbf{Open source}
			\\\hhline{|=|=|=|=|=|=|}
			AutoQML~\cite{klau2023qml} & \gcm   & \gcm &  \gcm & \gcm & \rx \\ \hline
			AutoQML~\cite{ALTARESLOPEZ2024} & \rx & \gcm  &  \gcm & \rx & \gcm \\\hline
			AutoAnsatz & \rx  & \gcm & \gcm & \rx & \rx \\\hline
			AQMLator & \gcm & \gcm & \gcm & \gcm  & \gcm \\\hline
		\end{tabular}
	\end{table}

\subsection{Design principles}
\label{sec:design}

	There are several key ideas behind the development of AQMLator. First and
	foremost it is meant to be used by the non-QC and even non-ML practitioners.
	Ideally, after the installation the user should only provide the data and a
	task for which the model should be proposed. AQMLator will search for, and
	ultimately suggest a quantum machine learning model. We present the
	intended usage in the following code snippet.
\begin{lstlisting}[language=Python]
# Initialize the model finder.
classifier_finder: ModelFinder = ModelFinder(
	task_type=MLTaskType.CLASSIFICATION,
	features=cls_x,
	classes=cls_y,
	n_cores=1,  # Can be left blank.
	n_trials=n_trials,  # Can be left blank.
	n_seeds=n_seeds,  # Can be left blank.
	n_epochs=n_epochs,  # Can be left blank.
	device=dev,  # Will use default simulator if left blank.
)
# Find the model.
model_finder: QMLModel = classifier_finder.find_model()
\end{lstlisting}
	The ease of access is further improved by the detailed documentation
	\cite{AQMLatorDocs}, well-commented code and numerous examples of usage
	found in the unit tests distributed with the code.

	The simplicity of use doesn't impair our solution in any way. AQMLator
	covers a range of standard ML tasks (classification, linear regression,
	clustering) and a number of models (QNN, Quantum Kernels, Restricted Boltzmann Machines). It was also designed with the extendability and customization in mind. For example, default distribution of AQMLator suggests two data embedding methods: \texttt{AngleEmbedding} or 
	\texttt{AmplitudeEmbedding}. One can see that in the one of configuration
	\texttt{dict}s of the \texttt{tuner.py} module.
\begin{lstlisting}[language=Python]
data_embeddings: Dict[str, Dict[str, Any]] = {
	"ANGLE": {"constructor": AngleEmbedding, "kwargs": {}, "fixed_kwargs": {}},
	"AMPLITUDE": {
		"constructor": AmplitudeEmbedding,
		"kwargs": {},
		"fixed_kwargs": {"pad_with": 0, "normalize": True},
	},
}
\end{lstlisting}

	\noindent Extending the \texttt{data\_embeddings dict} with different,
	\texttt{pennylane}-compatible embedding methods, will automatically
	increase the available search-space of the \texttt{ModelFinder}. Additional
	out-of-the-box QML model layers and data embedding methods are described in
	the Pennylane documentation~\cite{PennyLaneDocs}. Similar extensibility
	also applies to the other parts of the framework (models, layers,
	optimizers, ML tasks), yet it may require competences in either ML or QC.

	AQMLator implements the notion of limited quantum resources budget and aims
	to find such QML model that will require the least number of quantum device
	calls during the training on actual device, while at least maintaining the
	evaluation metric threshold specified by the user. By default AQMLator uses 
	mean accuracy and silhouette scores for supervised and unsupervised models
	respectively. Both are standard ML models evaluation methods. The user
	can easily introduce their own evaluation functions either globally, by
	modifying the \texttt{QMLModel.score} method or for each model separately.
	
	AQMLator is a framework developed in \texttt{Python} 3.11 using standard
	(Q)ML libraries, such as \texttt{PennyLane}, \texttt{sklearn}
	(\texttt{SciKit-Learn}), \texttt{PyTorch} and \texttt{Optuna} with a
	\texttt{PostgreSQL} database. \texttt{PennyLane} provided the basic
	building blocks for our supervised learning models. We used \texttt{sklearn}
	\texttt{mixins} to provide a standard interface for our models.
	\texttt{PennyLane} allows one to import its models as \texttt{PyTorch}
	layers. This makes introduction of the AQMLator-found models into the
	current ML pipelines seamless. \texttt{Optuna} is the heart of our
	framework, as it's used for model selection, quantum architecture search
	and hyperparameter optimizaiton. Finally, with \texttt{optuna-dashboard}
	we are also able to interactively review experiments' results stored in the
	\texttt{PostreSQL} database. Our unsupervised learning module is an
	extension of the implementation provided by QBM4EO project~\cite{QBM4EO}.
	AQMLator is a fully-documented, unit-tested, open-source software available
	via PyPi~\cite{AQMLatorPyPI}. 

\subsection{Quantum Architecture Search}
	
	AQMLator treats quantum architecture search as a hyperparameter
	optimizaiton task. By doing so, we can use \texttt{Optuna}
	to find both model and its architecture. During the \texttt{Optuna}
	optimization process, we first sample the number of layers that the model
	will have.
\begin{lstlisting}[language=Python]
kwargs: Dict[str, Any] = {
	"wires": len(self._x[0]),
	"n_layers": trial.suggest_int(
		"n_layers" + self._optuna_postfix,
		self._models_dict[model_type]["n_layers"][0],
		self._models_dict[model_type]["n_layers"][1],
	),
	"n_epochs": self._n_epochs,
	"accuracy_threshold": self._minimal_accuracy,
}
\end{lstlisting}
	The layers number's bound, both maximal and minimal, can be customized
	for each model in one of the tuner's customization \texttt{dict}s. For
	instance, in the default AQMLator distribution, we can see that our
	\texttt{QNNBinaryClassifier} has 1--3 layers, while our
	\texttt{QuantumKernelBinaryClassifier} can have 3--5 layers.
\begin{lstlisting}[language=Python]
binary_classifiers: Dict[str, Dict[str, Any]] = {
	"QNN": {
		"constructor": QNNBinaryClassifier,
		"kwargs": {
			"batch_size": (15, 25),  # Might need to be data size-dependent instead.
		},
		"fixed_kwargs": {},
		"n_layers": (1, 3),
	},
	"QEK": {
		"constructor": QuantumKernelBinaryClassifier,
		"kwargs": {},
		"fixed_kwargs": {},
		"n_layers": (3, 5),
	},
}
	\end{lstlisting}
	With the layers' number set, AQMLator uses \texttt{Optuna} to propose
	the data embedding method and the following layers of the model. 
\begin{lstlisting}[language=Python]
def _suggest_supervised_model_kwargs(
	self, trial: optuna.trial.Trial, model_type: str
) -> Dict[str, Any]:

	<...>
	
	self._suggest_embedding(trial, kwargs)
	self._suggest_layers(trial, kwargs)
	
	<...>
\end{lstlisting}
	AQMLator samples \texttt{Pennylane} data embedding methods and QML model
	layers, or rather their constructors, as categorical parameters. Additional
	parameters, if present, can be specified in one of the configuration \texttt{dict}s in \texttt{tuner.py}.
\begin{lstlisting}[language=Python]
def _suggest_layers(
	self, trial: optuna.trial.Trial, kwargs: Dict[str, Any]
) -> None:

	<redacted docstrings>
	
	layers: List[Type[qml.operation.Operation]] = []

	for i in range(kwargs["n_layers"]):
		layer_type: str = trial.suggest_categorical(
			f"layer_{i}" + self._optuna_postfix, list(layer_types)
		)
		layers.append(layer_types[layer_type]["constructor"])

	kwargs["layers"] = layers
	kwargs.pop("n_layers")  # No longer needed.
\end{lstlisting}
	After layers sampling is done, the model creation is possible. It requires
	calling the constructors in the same order they were sampled. We repeat
	this process a user-specified number of times. The model that scored best,
	that is met the quality metric threshold and minimized the number of quantum
	device calls, is returned by the end of the search.

\subsection{Hyperparameters optimization}
\label{sec:HPO}
	
	Hyperparameters optimization with AQMLator is a two-part process. Main part
	of the optimization	takes place during the model search. Analogously to the
	data embedding method and model layers, a number of additional parameters
	is suggested by \texttt{Optuna}. Consider the following method of the \texttt{ModelFinder}.\\
\begin{lstlisting}[language=Python]
def _suggest_unsupervised_model_kwargs(
	self, trial: optuna.trial.Trial, model_type: str
) -> Dict[str, Any]:
	lbae_input_shape: Tuple[int] = (1,) + np.array(self._x[0]).shape
	lbae_input_size: int = prod(lbae_input_shape)
	
	kwargs: Dict[str, Any] = {
		"lbae_input_shape": lbae_input_shape,
		"lbae_out_channels": trial.suggest_int(
			name="lbae_out_channels",
			low=floor(sqrt(lbae_input_size)),
			high=ceil(0.75 * lbae_input_size),
		),
	}
	
	kwargs["rbm_n_visible_neurons"] = kwargs["lbae_out_channels"]
	
	kwargs["rbm_n_hidden_neurons"] = trial.suggest_int(
		name="rbm_n_hidden_neurons",
		low=floor(sqrt(kwargs["lbae_out_channels"])),
		high=ceil(0.75 * kwargs["lbae_out_channels"]),
	)
	
	kwargs_data: Dict[str, Any] = self._models_dict[model_type]["kwargs"]
	
	kwargs["lbae_n_layers"] = trial.suggest_int(
		"lbae_n_layers" + self._optuna_postfix,
		kwargs_data["lbae_n_layers"][0],
		kwargs_data["lbae_n_layers"][1],
	)
	
	kwargs["fireing_threshold"] = trial.suggest_float(
		"fireing_threshold" + self._optuna_postfix,
		kwargs_data["fireing_threshold"][0],
		kwargs_data["fireing_threshold"][1],
	)
	
	kwargs["n_epochs"] = self._n_epochs
	
	return kwargs
\end{lstlisting}	
	One can see that \texttt{lbae\_out\_channels}, 
	\texttt{rbm\_n\_hidden\_neurons}, \texttt{lbae\_n\_layers},
	\texttt{n\_epochs}, all hyperparameters of the unsupervised model, are
	selected with \texttt{Optuna}. In the case of supervised learning
	models the number of layers is the most important hyperparameter.
	
	The second part of HPO using AQMLator concerns model training. If the user
	would like to ensure that the default training method --- \texttt{PennyLane}
	\texttt{GradientDescentOptimizer} --- is the optimal one, they can use our
	\texttt{HyperparameterTuner} to compare the training performance
	using
	different optimizers. This is done by instantiating the tuner
	and calling its only public method, as presented on the code snipped
	below. \\ 
\begin{lstlisting}[language=Python]
tuner: HyperparameterTuner = HyperparameterTuner(
	x,
	y,
	classifier,  # Or any other QMLModel.
	n_seeds=n_seeds,
	n_trials=n_trials,
)
tuner.find_hyperparameters()
\end{lstlisting}	

\subsection{Experiments revision with optuna-dashboard}
\label{sec:exp}
\begin{figure}[htbp] 
	\centering
	\includegraphics[width=1\textwidth]{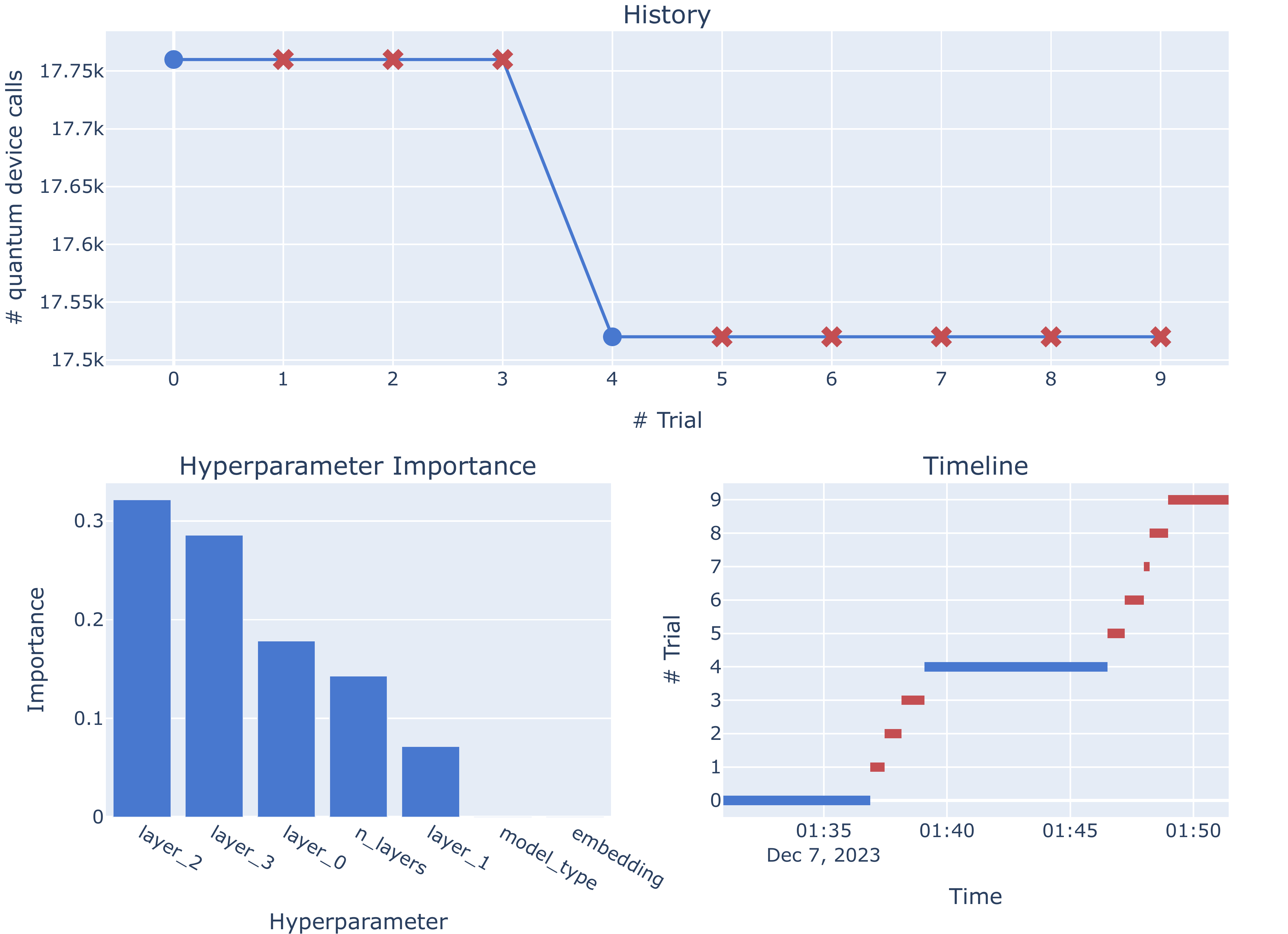}
	\caption{An example results of model finding using AQMLator. Do note
	that additional information, specifically values of the hyperparameters
	for a given trial, are accessible via interaction with the plot in
	the \texttt{optuna-dashboard}.}
	\label{fig:dashboard_model_finder}
\end{figure}

	Once model finding or hyperparameter tuning experiments are done, their
	results can be interactively reviewed using \texttt{optuna-dashboard}. 
	For that to happen, however, the user has to specify one of the available
	types of databases as the \texttt{Optuna} storage. We present an example
	of how \texttt{optuna-dashboard} presents the experiment results in the
	Figure~\ref{fig:dashboard_model_finder}. Notice that
	\texttt{optuna-dashboard} also allows one to access additional
	info about each data point over plot on mouse hover. This feature is not
	shown in the figure.
	
\section{Conclusions and future work}
	In this paper, we present AQMLator --- an auto, quantum machine learning
	platform designed for users without knowledge in quantum computing and very
	limited experience in machine learning. We argue that the minimal
	requirements presented before the user will allow a broad audience to
	include QML techniques in their ML pipelines, thus filling the gap in the
	field of accessible auto quantum machine learning platforms.
	
	Our approach is open source, can be run on both simulators and real
	quantum devices via \texttt{qiskit}. It fully automatically suggests
	the QML model given only the data and task from the user. The platform was
	built using standard (Q)ML libraries, allowing a seamless introduction
	of the resultant models to the existing (Q)ML pipelines. It's also fully
	documented, making it easy to use, extend and customize to one's needs.
	
	Although complete in its current form, there's a lot of ways AQMLator
	could be improved. A natural course of action would be to introduce
	additional layers, data embedding methods and models to the default
	version of the framework, so that the users won't have to specify them
	themselves. Another approach would be to introduce advanced
	reinforced learning-based techniques of quantum architecture search to our
	framework, so that inspection of the whole available circuit space would be
	done in a more	complete fashion. We could also further simplify the
	process of adding additional evaluation metrics to AQMLator and enable
	mutliobjective optimization native to \texttt{Optuna}. This way the user
	could directly specify the desired cost-to-quality ratio.

\section*{Acknowledgment}
	The authors would like to thank Etos sp. z o.o. and the QBM4EO team
	for providing the QBM4EO code.
	We	acknowledge the financial support by the EuroHPC PL project co-financed
	by EU within the Smart Growth Operational Programme (contract no.
	POIR.04.02.00-00-D014/20-00). We gratefully acknowledge the funding support
	by program ``Excellence Initiative --- Research University'' for the AGH
	University of Krakow as well as the ARTIQ project ARTIQ/0004/2021.

\bibliographystyle{plain}
\bibliography{bibliography.bib}
\end{document}